\magnification = \magstep1
\vsize = 22 true cm
\hsize = 16 true cm
\baselineskip = 24 true pt
\centerline {\bf {PATH INTEGRAL FORMULATION OF THE CONFORMAL }} 
\centerline {\bf {WESS-ZUMINO-WITTEN $\rightarrow$ TODA REDUCTIONS}} 
\bigskip
\bigskip
\centerline {L. O'Raifeartaigh and V. V. Sreedhar} 
\centerline {School of Theoretical Physics}
\centerline {Dublin Institute for Advanced Studies} 
\centerline {10, Burlington Road, Dublin 4}
\centerline {Ireland}
\bigskip
\bigskip
\centerline {\bf {Abstract}}
The phase space path integral Wess-Zumino-Witten $\rightarrow$ Toda reductions 
are formulated in a manifestly conformally invariant way. For this purpose, the 
method of  Batalin, Fradkin, and Vilkovisky, adapted to conformal  
field theories, with chiral constraints, on compact two dimensional 
Riemannian manifolds, is used. It is shown that the  
zero modes of the Lagrange multipliers produce the Toda potential and the  
gradients produce the WZW anomaly. This anomaly is crucial 
for proving the Fradkin-Vilkovisky theorem concerning gauge     
invariance. 
\vfill
\hfill DIAS-STP-98-05
\vfil\eject
\bigskip
\centerline {\bf I. INTRODUCTION AND THE STATEMENT OF THE PROBLEM}
\bigskip
It is well known that classical Wess-Zumino-Witten (WZW) models based on simple 
finite dimensional groups can be reduced to Toda theories by 
imposing linear first class constraints on the WZW currents [1].  
The quantised version of the reduction process was also considered  
earlier, but mainly within the framework of canonical quantisation [2].
The elegance of the classical reduction process 
suggests, however, that the most 
natural framework for quantisation is through the path integral. In a recent  
paper [3], we presented the path integral formulation of the 
simplest of such reductions, namely, the reduction of the $sl(2, R)$ WZW 
model to the Liouville theory. In that paper we stressed the  
importance of the zero modes in proving gauge invariance and in  
producing the WZW anomaly and the Liouville  potential.  
The zero modes occur because we must work on a compact manifold since  
otherwise it is not possible to choose 
configurations for which the kinetic term and the exponential potential  
are both finite.  
In the present paper we present a generalisation of 
these results to the reduction of WZW theories to Toda theories.  

The generalisation from the Liouville to the Toda case is not trivial for a  
number of reasons. First, the usual off-diagonal parameters in the Gauss 
decomposition for the group valued fields are not 
the natural fields from the path integral point of view because the 
Lagrangian density is not quadratic in these fields. In addition, with 
respect to the `improved' Virasoro generators which are necessary to 
implement the constraints, these fields are not primary.    
However, they can be converted into primary fields for which the 
Lagrangian density is bilinear and remains local.  The new fields have the  
further property that the first class constraints can be expressed as linear  
conditions on their conjugate momenta.
Second, the separation of the Lagrange multipliers into their zero modes 
and their orthogonal complements requires the    
introduction of auxiliary fields. The auxiliary fields are only needed to 
define orthogonality in a  
conformally invariant manner and 
can be eliminated once the  
zero modes are separated.    
Third, it turns out that the zero modes occur only at grade one.   
Finally, unlike the case in the $sl(2, R)$ WZW $\rightarrow$ Liouville 
reduction, there are  ghost fields of non-zero conformal weights  
and these  give a non-trivial  
contribution to the Polyakov term 
and hence to the reduced Virasoro centre. 

Another  interesting feature is that the operations of reducing and quantising 
do not commute in the sense that they lead to Toda theories with different  
coupling constants depending on the order in which the operations are 
performed. The coupling constants are $k$ and $k-\gamma$ respectively,   
where $k$ is the WZW  
coupling constant and $\gamma$ is the dual 
Coxeter number of the group.  
As might be expected, the shift in the coupling constant 
originates in the WZW anomaly.  

As in the WZW $\rightarrow$ Liouville reduction, we use the  
Batalin-Fradkin-Vilkovisky (BFV) generalisation of the Faddeev-Popov 
formalism [4]. 
This is because the BFV formalism allows us to use  
the WZW gauge in which the Lagrange multipliers are 
zero (the analogue of the temporal gauge in QED). 
Since the manifold is compact, and the constraints are chiral, we  
use the modification of the standard BFV formalism 
which was introduced in [3].  
As in the Liouville case, the integration over the zero modes produces the    
exponential potential and the gauge variant gradient parts produce the WZW  
anomaly.  
The anomaly is crucial for proving the 
Fradkin-Vilkovisky theorem regarding the gauge independence of the 
reduction.  

The paper is organised as follows. In section {\bf II}, we briefly recall the 
basics of simple Lie groups and define the principal $sl(2, R)$ embedding. In  
section {\bf III}, we review the classical WZW $\rightarrow$ Toda reductions, 
based 
on simple Lie groups with principal $sl(2, R)$ embeddings, to Toda theories.
In section {\bf IV}, we summarise the basics of the BFV formalism. Section  
{\bf V} contains the main results of this paper and establishes the quantum 
WZW $\rightarrow$ Toda reductions in a gauge independent manner, 
by using a modification of the BFV 
path integral that takes into account the chiral nature of the constraints 
and the compactness of the manifold. In section {\bf VI}, we examine the 
results of the previous section in two special gauges which highlight 
the general results.  
In section {\bf VII} we present our conclusions.  
\bigskip
\centerline {\bf II. SIMPLE LIE GROUPS AND THE PRINCIPAL sl(2, R) EMBEDDING}
\bigskip
In this section we briefly recall the necessary properties of simple 
finite dimensional Lie groups $G$ [5]. The standard Cartan-Weyl basis for  
the Lie algebra $\cal G$ is   
$$[H_i, H_j] = 0,~~~~~~
[H_i, E_\alpha ] = \alpha_i E_\alpha ,
~~~~~~ 1\leq i, j \leq l
\eqno(2.1a)$$ 
$$\eqalign {[E_\alpha , E_\beta ] &= N_{\alpha\beta}E_{\alpha +\beta},
~~{\hbox {if}}~~\alpha + \beta \in \Delta\cr
&= {2\alpha\cdot H\over \alpha^2}
\delta_{\alpha, -\beta} 
\cr }\eqno(2.1b)$$ 
where the $\alpha_i$ are components of root vectors, $\Delta$ is the lattice 
of root vectors, and the
$N_{\alpha\beta}$ are constants.  
In the orthonormal basis for the Cartan subalgebra, 
$$ Tr(H_iH_j ) = \delta_{ij},~~~
Tr(E_\alpha , E_\beta ) = {2\over\alpha^2}
\delta_{\alpha , -\beta}\eqno(2.2)$$
The second equation in (2.2), which is actually a consequence of the first one, 
and (2.1), fix the normalisation of $E_\alpha$. 
The Lie algebra 
$\cal G$ 
is said to be simply-laced if all the roots $\alpha$ have the same squared 
length $\alpha^2 = \alpha_i\alpha_i$ (as is the case for the 
$A, D$ and $E$ series) and non-simply laced otherwise.  
The real span of the Cartan-Weyl basis yields a 
maximally non-compact real form $\cal G_R$ of the Lie algebra $\cal G$.  
These are the Lie algebras in which we are interested.  
For the classical 
Lie algebras $A_n$, $B_n$, $C_n$ and $D_n$, these forms are given by the 
real Lie algebras $sl(n, R), so(p+1, p, R), sp(2n, R)$, and $so(p, p, R)$. 

Since the number of root pairs $\pm\alpha$, in general, 
exceeds the rank $l$, it is convenient to choose a set of roots $\alpha^{(s)}, 
1\leq s \leq l$, which constitute a basis for the $l$-dimensional space 
of roots. This basis can be chosen in such a way that an arbitrary root 
$\alpha$ can always be expressed as 
$$\alpha = \sum_{s = 1}^l n_s\alpha^{(s)}\eqno(2.3)$$ 
where each $n_s\in Z$ and either $n_s \geq 0$ or $n_s\leq 0$. The two  
cases correspond to $\alpha$ being a positive or a negative root respectively.  
The $\alpha^{(s)}$ are said to constitute a basis of simple roots which 
we will denote be $\Delta_s$, and the subspaces of positive and negative roots 
have the obvious notation $\Delta^+$ and $\Delta^-$. For the simple roots 
$$[E_{\alpha^{(s)}} , E_{-\alpha^{(r)}}] = 
{2\alpha^{(s)}\cdot H\over \mid{\alpha^{(s)}}\mid^2}
\delta_{rs}\eqno(2.4)$$ 
The principal $sl(2, R)$ embedding in $\cal {G_R}$  
is defined by choosing an element $M_0$ in the  
Cartan subalgebra, with respect to which all the simple roots have grade one.
This element is given uniquely by  
$$ M_0 = \rho\cdot H,~~~{\hbox{where}}~~~
\rho = \sum_{s =1}^l
\mu^{(s)}~~~= {1\over 2}\sum_{\alpha\in\Delta^+}\alpha\eqno(2.5)$$
where $\mu^{(s)}$ are the fundamental weights.  
The other components of the principal $sl(2, R)$ embedding in $\cal {G_R}$ are 
got by taking suitable linear combinations $M_\pm$ of $E_{\pm\alpha^{(s)}}$.   
$$M_+ = \sum_s p_s E_{\alpha^{(s)}}~~~{\hbox{and}}~~~
M_- = \sum_s q_sE_{-\alpha^{(s)}}\eqno(2.6)$$
where the sum extends only over the simple roots because 
the $M_\pm$ are step operators with grade $\pm 1$ with respect to $M_0$.   
The requirement that $\{M_0, M_\pm\}$ constitute an $sl(2, R)$ subalgebra 
constrains the combination $p_sq_s$ to be of the form  
$$ p_sq_s = \alpha^{(s)} (K^{-1})_{sr}\rho^{(r)}~~~{\hbox{where}}~~~
K_{rs} = 2{\alpha^{(r)}\cdot\alpha^{(s)}\over \mid\alpha^{(s)}\mid^2}
\eqno(2.7)$$
$K$ being the Cartan matrix.  
In conjunction with (2.3) we see that $M_0$ defines an integer grading for 
the entire Lie algebra $\cal G_R$. We will now review the basics of the WZW  
models based on the real forms of simple, non-compact, groups $G_R$ and and   
use the principal $sl(2, R)$ embedding defined above to reduce them classically 
to Toda theories.  
\vfil\eject
\centerline {\bf III. THE CLASSICAL WZW $\rightarrow$ TODA REDUCTIONS}
\bigskip
The WZW model based on a group $G_R$ is defined on a two dimensional 
compact manifold $\partial\Sigma$ 
by the Action [6]    
$$S = k\int_{\partial\Sigma} Tr~(g^{-1}d g)
\cdot (g^{-1}d g) - {2k\over 3}\int_\Sigma Tr~ 
(g^{-1}d g)\wedge (g^{-1} d g)\wedge (g^{-1}d g)
\eqno(3.1) $$
Here $g\in G_R$. 
The two dimensional manifold is parametrised by the light-cone coordinates 
$z_r$ and $z_l$ defined by 
$$z_r = {z_0 + z_1 \over 2},~~~z_l = {z_0 - z_1\over 2}\eqno(3.2)$$
respectively. 
The Action is invariant under 
$$g\rightarrow gu(z_r ),~~~ g\rightarrow v(z_l )g\eqno(3.3) $$
where $u(z_r), v(z_l) \in G_R$.  
The conserved Noether currents which generate the above transformations, 
$$J_r = -(\partial_r g)g^{-1}, ~~~~J_l = g^{-1}(\partial_l g)\eqno(3.4)$$
take their values in the Lie algebra $\cal G_R$. As is well-known, the 
components $J_r^a$ of the currents satisfy the classical version of the 
Kac-Moody algebra, given in terms of the Poisson brackets of the currents by 
$$\{J_r^a (z), J_r^b(z')\} = if^{ab}_cJ_r^c\delta (z - z') + k\delta^{ab}
\delta (z - z')\eqno(3.5)$$
A similar equation also holds for the components of the left currents.  

Since we are interested only in the maximally non-compact real form of the 
Lie algebra, the group element $g$ admits a unique, local, Gauss decomposition 
of the form 
$$g = ABC\eqno(3.6)$$
where
$$A = e^{\sum\hat a^\alpha E_\alpha},~~~B = e^{\phi\cdot H},~~~
C = e^{\sum\hat a^{-\alpha}E_{-\alpha}}~~~{\hbox{where}}~~~
\alpha\in\Delta^+\eqno(3.7)$$ 
In the above equation, $A$ and $C$ are nilpotent subgroups each with 
dimension $({\hbox{dim}}~G - l)/2$, and $B$ is the maximal abelian subgroup. 
This property makes the WZW models based on real, non-compact, groups, 
the natural generalisations of the $SL(2, R)$ model studied in [3]. 
The fields $\hat a^\alpha$, $\hat a^{-\alpha}$, and $\phi^i$ parametrise 
the group manifold. 
As is well-known, the Gauss decomposition is not valid globally. This issue 
has been dealt with in detail in [7]. For simplicity, we restrict  
our present considerations to the coordinate patch that contains the 
identity. Similar results hold for the other patches. The above decomposition 
is very useful for setting up the Hamiltonian formalism to which we now pass.   

The Polyakov-Wiegmann factorisation formula [8] for the WZW model 
states that
$$S(XY) = S(X) + S(Y) + \int d^2z~Tr[(X^{-1}\partial_lX)(\partial_rY)Y^{-1}]
\eqno(3.8)$$
where $X\in G_1$ and  $Y\in G_2$, $G_1$ and $G_2$ being arbitrary simple Lie  
groups. Using this formula recursively, and from the nilpotency of $A$ and $C$,
we find 
$$ S(ABC) =  S(B)  
+\int d^2z~Tr[(A^{-1}\partial_lA)B(\partial_rC)C^{-1}B^{-1}]\eqno(3.9)$$
Substituting for $A, B$ and $C$ from (3.7) in the above equation and 
evaluating the traces gives 
$$S = {k\over 2}\int_{\partial\Sigma }d^2z~\Bigl[(\partial_l\phi^i)
(\partial_r\phi^i) + {4\over\alpha^2}V_\alpha 
U_{-\alpha '} e^{-\alpha\cdot\phi} (\partial_l\hat a^\alpha )
(\partial_r \hat a^{-\alpha '})
\Bigr]\eqno(3.10)$$ 
where $U$ and $V$ are defined by the left and right currents of the 
nilpotent subgroups $A$ and $C$ through the relations
$$A^{-1}d A = V_\alpha (\hat a^\alpha )d \hat a^\alpha = V_\alpha^\beta (
\hat a^\alpha )E_\beta d\hat a^\alpha\eqno(3.11a)$$
$$(dC)C^{-1} = U_{-\alpha} (\hat a^{-\alpha} )d \hat a^{-\alpha} = 
U_{-\alpha}^{-\beta} (
\hat a^{-\alpha} )E_{-\beta} d\hat a^{-\alpha}\eqno(3.11b)$$
It is clear that because $U$ and $V$ are functions of $\hat a^{-\alpha}$ and 
$\hat a^\alpha$ respectively, the Action (3.10) is not quadratic in these 
fields. Since we would finally like to integrate out these fields,  
the off-diagonal parameters in the Gauss 
decomposition of the group valued fields are not the natural ones to use as  
fields. However, the functions $U$ and $V$ can be used as kernels to define new 
fields $a^\alpha$ and $a^{-\alpha}$ expressed in terms of the old fields 
$\hat a^\alpha$ and $\hat a^{-\alpha}$ through the integral equations 
$$a^\alpha =  \hat a^\alpha + \int \sum_{\beta < \alpha} V_\beta^\alpha 
(\hat a^\alpha )d\hat a^\beta , 
~~~{\hbox {and}}~~~ a^{-\alpha}  
= \hat a^{-\alpha} + \int \sum_{\beta < \alpha} U_{-\beta}^{-\alpha}
(\hat a^{-\alpha} )d\hat a^{-\beta}  
\eqno(3.12)$$
All the results will be completely independent of these kernels. 
In terms of the new fields the Action takes the simple form 
$$S = {k\over 2}\int_{\partial\Sigma }d^2z~\Bigl[(\partial_l\phi^i)
(\partial_r\phi^i) + {4\over\alpha^2}
e^{-\alpha\cdot\phi} (\partial_la^\alpha )(\partial_r a^{-\alpha })
\Bigr]\eqno(3.13)$$ 
Notice that the Lagrangian density in (3.13) is quadratic in the new 
fields $a^{\pm\alpha}$. Moreover, the Lagrangian density remains local 
although the transformation in (3.12) is not. Furthermore, the transformation 
is idempotent and hence the Jacobian of the transformation is unity. Thus 
there will be no change in the standard symplectic measure used to define the 
phase space path integral in section {\bf V}. 
The momenta canonically conjugate to $a^\alpha , \phi^i , a^{-\alpha}$
respectively are  
$$\pi_\alpha  = 
 {2k\over\alpha^2}(\partial_r a^{-\alpha} )e^{-\alpha\cdot\phi},~~~ 
\pi_{-\alpha} =   
 {2k\over\alpha^2}(\partial_l a^\alpha )e^{-\alpha\cdot\phi },~~~ 
\pi_i  = 
k\partial_0\phi^i \eqno(3.14)$$ 
The canonical Hamiltonian density $H$ is 
$$ H = {1\over 2k}\pi_i^2 +  {k\over 2}(\phi^i )^{'2} + 
{\alpha^2\over 2k}\pi_\alpha\pi_{-\alpha} e^{\alpha\cdot\phi} +
\pi_\alpha
(a^\alpha )' - \pi_{-\alpha}(a^{-\alpha})' \eqno(3.15)$$
The left and right conserved currents are given by
$$J_l = g^{-1}\partial_l g = C^{-1}[e^{-\alpha\cdot\phi}(\partial_l a^\alpha )
E_\alpha + (\partial_l\phi^i )H_i + (\partial_la^{-\alpha})E_{-\alpha}]C
\eqno(3.16a)$$
$$J_r = -(\partial_r g)g^{-1} = - A[(\partial_ra^\alpha )E_\alpha + 
(\partial_r\phi_i)H_i + (\partial_ra^{-\alpha})BCE_{-\alpha}C^{-1}B^{-1} ]
A^{-1}\eqno(3.16b)$$ 
It is straightforward to check that, in terms of the currents, the Hamiltonian 
density takes the Sugawara form {\it viz.}
$$H = {\cal T}_r + {\cal T}_l~~~{\hbox{where}}~~~{\cal T}_r 
= {1\over 2}Tr~[J_r^2]
~~{\hbox{and}}~~{\cal T}_l = {1\over 2}Tr~[J_l^2]\eqno(3.17)$$
The currents may also be expressed completely in terms of the phase space 
variables $a^\alpha ,\phi^i , a^{-\alpha}$ and their conjugate momenta using 
the relations in (3.14).  Further, by using canonical Poisson brackets  
for the phase space variables {\it viz.}
$$\{a^\alpha (z) ,\pi_\beta (z')\} = 
\{a^{-\alpha} (z) ,\pi_{-\beta} (z')\} = \delta^\alpha_\beta
\delta (z-z');
~~\{\phi^i (z),\pi_j (z')\} = \delta^i_j\delta (z -z') 
\eqno(3.18)$$ 
the rest being zero, we can check explicitly that the currents satisfy two 
independent copies of the standard Kac-Moody algebra (3.5). This is a further 
proof of the fact that the measure for the phase space path integral in 
terms of the new fields $a^{\pm\alpha}$ is the standard symplectic measure.   

The constraints we want to impose are 
$$ \Phi_\alpha \equiv J^r_\alpha - M_\alpha = 0~~{\hbox{where}}~~M_\alpha 
\neq 0~~ {\hbox{iff}}~~\alpha\in\Delta_s\eqno(3.19a)$$ 
and 
$$ \Phi_{-\alpha} \equiv J^l_{-\alpha} - M_{-\alpha} = 0~~{\hbox{where}}~~
M_{-\alpha}  
\neq 0~~ {\hbox{iff}}~~-\alpha\in\Delta_s\eqno(3.19b)$$ 
where $M_\alpha$ and $M_{-\alpha}$ are the components of the step operators 
of the principal $sl(2, R)$ embedding, 
$$ M_\alpha = Tr~(E_{-\alpha}M_+)~~~{\hbox{and}}~~~M_{-\alpha} = 
Tr~(E_\alpha M_-)\eqno(3.19c)$$
It may be worth mentioning in passing here that, for the purposes of this 
paper, the above requirement is not very strict and $M_\alpha$ and 
$M_{-\alpha}$ can be allowed to be completely arbitrary, but non-zero, 
constants. That they are the components of $M_\pm$ is necessary only 
to ensure that the reduced Kac-Moody algebra is a W-algebra {\it i.e.} 
has a primary basis. 

The virtue of using the new variables $a^{\pm\alpha}$ is that the constraints 
(3.19) can be directly expressed in terms of their conjugate momenta.  
$$ \Phi_\alpha \equiv \pi_\alpha - {2k\over\alpha^2}M_\alpha =
 0~~{\hbox{where}}~~M_\alpha 
\neq 0~~ {\hbox{iff}}~~\alpha\in\Delta_s\eqno(3.20a)$$ 
and 
$$ \Phi_{-\alpha} \equiv \pi_{-\alpha} - {2k\over\alpha^2}M_{-\alpha}
 = 0~~{\hbox{where}}~~
M_{-\alpha}  
\neq 0~~ {\hbox{iff}}~~-\alpha\in\Delta_s\eqno(3.20b)$$ 
This is possible because the constraints (3.19) reduce the  relationship 
between the currents and the momenta $\pi_\alpha$ and $\pi_{-\alpha}$, 
which in general is quite complicated, to a simple linear relation.  
Upon imposing the constraints (3.20) on the classical Hamiltonian density 
(3.15) of the WZW model, we get, apart from boundary terms, 
$$ H = {1\over 2k}\pi_i^2 +  {k\over 2}(\phi^i )^{'2} + 
k\Lambda_\alpha e^{\alpha\cdot\phi} 
 \eqno(3.21)$$
where 
$$\Lambda_\alpha = {2\over\alpha^2}M_\alpha M_{-\alpha}\eqno(3.22)$$
This is easily recognised as the expression for the Hamiltonian density of the  
classical Toda theory. Although the two sets of constraints (3.19) and 
(3.20) are completely equivalent physically in the above sense, it is 
considerably simpler to work in terms of the latter set because, unlike 
the Poisson bracket of two current components,  the Poisson 
bracket of two momenta is 
strictly zero. This fact implies that the BRS charge for the reduction, 
to be defined in the next section,  
does not have terms which involve higher powers of the ghost fields.   

The constraints in (3.20) set the grade one momenta 
$\pi_{\pm\alpha^{(s)}}$ equal to non-zero 
constants. As is well-known, this is not consistent with 
the conformal invariance, defined by the two Sugawara Virasoro operators  
in (3.17), because the momenta, like the corresponding currents  
$J^r_{-\alpha^{(s)}}$ and $J^l_{\alpha^{(s)}}$ have conformal dimension one.   
Hence, the Virasoro generators in (3.17) are replaced by  
the `improved' generators $T_r$ and $T_l$ defined by 
$$T_r = {\cal T}_r - {2\over \alpha^2}\partial_r J_r^0
~~~{\hbox {and}}~~~ 
T_l = {\cal T}_l - {2\over \alpha^2}\partial_l J_l^0\eqno(3.23)$$
where 
$$J_r^0 = Tr~[M_0J_r]~~~{\hbox{and}}~~~J_l^0 = Tr~[M_0J_l]\eqno(3.24)$$
As will be seen in section {\bf V}, this improvement may be implemented  
by coupling the currents to a fixed background metric in a specific, 
non-minimal, way. At this stage another advantage of using the new fields  
$a^{\pm\alpha}$ becomes clear. With respect to the  
`improved' Virasoros defined above, the new fields $a^{\pm\alpha}$ and hence 
their conjugate momenta $\pi_{\pm\alpha}$ are primary   
fields, unlike the old fields which do not have specific conformal 
transformation properties. With respect to the conformal group generated  
by the `improved' Virasoros, the currents we wish to constrain, namely,   
$J^r_{-\alpha}$ and $J^l_\alpha$, or equivalently the momenta $\pi_{\pm\alpha}$,
have the following conformal weights,
$$\omega (J^r_{-\alpha}) =  \omega (\pi_\alpha ) = (0, 1-m_\alpha ),~~~
\omega (J^l_\alpha ) = \omega (\pi_{-\alpha}) = (1-m_\alpha , 0)\eqno(3.25)$$
where the positive integer $m_\alpha$ is defined by 
$m_\alpha = \rho\cdot\alpha$. 
It follows that for simple roots ($\alpha\in\Delta_s$) which have grade 
$m_{\alpha^{(s)}}$ equal to one, the above currents and momenta are conformal 
scalars. The constraints in (3.19, 3.20) are, therefore, compatible with this   
conformal group.  

The currents $J_r^\alpha$ and $J_l^{-\alpha}$  
now have conformal weights $(0, 1+m_\alpha )$ and $(1+m_\alpha , 0)$
respectively.  
The phase space variables $a^\alpha $ and $a^{-\alpha}$ become primary fields 
of conformal weights $(0, m_\alpha )$ and $(m_\alpha , 0)$ respectively, 
the field $\alpha\cdot\phi$
becomes a conformal connection, while $e^{\alpha\cdot\phi} $ becomes a  primary 
field of weight $(m_\alpha , m_\alpha )$. It follows that for
$\alpha\in\Delta_s$, it has a 
weight (1,1) {\it i.e.} it has the opposite conformal weight to the volume 
element $d^2z$ in the two dimensional space. This completes our discussion 
of the classical aspects of the WZW $\rightarrow$ Toda reduction.   
The classical Toda theory (3.21) can be quantised by any standard procedure 
[2] and its Virasoro centre is given by 
$$c(k) = {l\over 6} + {2\rho^2\over k}\Bigl\lbrack 1 
+ k\Bigr\rbrack^2\eqno(3.26)$$
where $l = {\hbox{dim}}H$. This expression for the Virasoro centre, obtained 
by first reducing and then quantising, may be contrasted with the one 
obtained by first quantising using the path integral formulation and then 
reducing.   
The rest of the paper deals with this issue.  
\bigskip
\centerline {\bf IV. THE GENERAL PATH INTEGRAL REDUCTION PROCEDURE}
\bigskip
In this section we first give a brief sketch of the BFV formalism. 
Let $p$ and $q$ be any 
set of canonically conjugate variables, $H$ the canonical Hamiltonian, and  
$$Z = \int d(pq)~ e^{-\int dxdt~[p\dot q - H(p, q)]},\eqno(4.1)$$ 
the phase space path integral which is to be reduced by a set of 
first class constraints $\Phi (q, p)$. Let $A$ be a set of Lagrange 
multipliers, $B$ their canonically conjugate momenta, and $b, \bar c$ 
and $c, \bar b$ be conjugate ghost pairs. Then define the nilpotent BRST 
charge $\Omega$ by  
$$\Omega = \int dx~ [c\Phi + bB] + \cdots ~~~{\hbox{where}}~~~ 
\{\Omega , \Omega \} = 0\eqno(4.2)$$ 
Here the dots refer to terms which involve higher powers of ghost fields 
(which actually 
do not occur in the present case). A  
minimal choice for the gauge fixing fermion $\bar \Psi$ is given by  
$$\bar\Psi = \bar c\chi + \bar b A\eqno(4.3)$$ 
where $\chi (p, q, A, B) $ is a set of gauge-fixing conditions.  
The BFV procedure then consists of inserting the reduction factor 
$$ F = \int d(ABb\bar b c\bar c )e^{-\int dxdt~[\bar b \dot c - \{\Omega , 
\bar \Psi \}]}\eqno(4.4)$$ 
into the path integral in (4.1). 
The Fradkin-Vilkovisky theorem states that the reduced path integral  
is independent of the choice of the gauge fixing fermion $\bar\Psi$. There  
are some exceptions to this theorem, mainly because of the Gribov problem [9].
However, for the example we are considering, the gauge group is nilpotent, and 
the path integral is shown to be independent of the gauge fixing conditions 
by explicit calculation.  
In the definition of the reduction factor above, it is not necessary 
to include the term $B\dot A + {\dot {\bar c}} b$ in the Action
because such a term can always be generated by letting  
$\chi \rightarrow \chi + \bar c\dot A$. 
The standard non-zero Poisson brackets for the variables  
$$ \{q (x), p (x')\} = \{A (x), B(x')\} = 
 \{b (x), \bar c (x')\} = 
 \{c (x), \bar b (x')\} = \delta (x - x') \eqno(4.5)$$  
imply that 
$$\{\Omega , \bar\Psi\} = (A\Phi + B\chi) - 
(\bar b b - \bar c [FP]c - \bar c [BFV]b)\eqno(4.6)$$ 
where the FP and BFV terms are defined by 
$$ \{\Phi (x), \chi (x')\} = [FP]\delta (x - x'), ~~
\{B(x), \chi (x')\} = [BFV]\delta (x - x') = -{\partial\chi\over \partial A}
\delta (x - x')\eqno(4.7)$$ 
Substituting for $\{\Omega , \bar\Psi\}$ in $F$ and doing the 
$\bar b b$ integrations yields 
$$ F = \int d(AB\bar cc) e^{\int dxdt ~[A\Phi + B\chi + \bar c\{[FP] + 
[BFV]\partial_t \}c ] }\eqno(4.8)$$
Assuming that $\chi$ is independent of the $B$-fields, as is usually the 
case, we may also integrate over them to get 
$$\eqalign{ F &= \int d(A\bar cc)\delta (\chi )~ 
e^{\int dxdt ~[A\Phi + \bar c\{[FP] + 
[BFV]\partial_t \}c ] }\cr  
 &= \int dA \delta (\chi ) det\Bigl([FP] + [BFV]\partial_t\Bigr)~
 e^{\int dxdt ~[A\Phi ]} } 
 \eqno(4.9)$$
Note that if $[BFV]$ is equal to zero, we recover the standard Faddeev-Popov 
insertion [10]. On the other hand, as is clear from (4.7), it is the presence 
of the $[BFV]$ term that allows the gauge fixing function $\chi$ to depend 
on the Lagrange multipliers. Thus the BFV path integral allows us to consider 
gauge fixing conditions that depend on the Lagrange multipliers on an equal 
footing with those which depend only on the phase space variables. In the 
next section we present a slightly modified BFV path integral for 
establishing the gauge independent quantum WZW $\rightarrow$ Toda reductions.  
\bigskip
\centerline{\bf V. THE QUANTUM WZW $\rightarrow$ TODA REDUCTIONS}
\bigskip
In order to set up the quantum WZW $\rightarrow$ Toda reductions through the  
path integral method, we first note that since the constraints are linear
in the momenta, it is natural to start    
with the unconstrained WZW {\it phase space} path integral, namely,  
$$ I (j) = \int d(\phi^i \pi_i a^{\pm\alpha}
\pi_{\pm\alpha} )~e^{-\int d^2z~[\pi_\alpha\dot a^\alpha + \pi_i\dot\phi^i +
 \pi_{-\alpha}\dot a^{-\alpha} -  H + j\cdot\phi ]}
\eqno(5.1)$$  
and to use the BFV formalism for the reduction. The standard symplectic measure 
$d(\phi^i\pi_i a^{\pm\alpha}\pi_{\pm\alpha} )$ used above 
is the correct phase space measure because an integration over the momenta 
with this measure produces the configuration space path integral with 
the correct (group invariant) measure {\it viz.}  
$d(e^{-\alpha\cdot\phi}a^{\pm\alpha}\phi^i)$.   
Here the external source, $j$, is 
attached only to $\phi $ since the other variables will be eliminated 
by the reduction.  

We now apply the BFV formalism to WZW $\rightarrow $ Toda 
reductions.  
The application will  
differ from the standard BFV formalism  
in two respects. First, since we are dealing with independent left 
handed and right handed constraints, it is convenient to replace the standard 
BFV formalism with a light-cone version. The light-cone version of the BFV 
formalism is introduced by replacing the space and time directions by the 
two branches of the light-cone 
parametrised by the light-cone coordinates $z_r$ and $z_l$, using a different 
branch as the time for each of the two constraints.  
However, since we will use the Euclidean version of the theory in the 
path integral, these coordinates 
actually get converted into holomorphic and anti-holomorphic 
coordinates. Thus  
all the fields will be functions of the holomorphic and anti-holomorphic 
variables and functions which depend only on one variable 
will be  holomorphic or anti-holomorphic functions. Second,   
for reasons already explained, we must work on a compact manifold and thus  
we need a curved space generalisation of the BFV formalism. The compactness 
of the manifold also entails the presence of zero modes for the Lagrange 
multiplier fields. As will be seen later, the need to define the orthogonality 
condition between the zero modes and the gauge variant modes in a conformally 
invariant manner requires us to introduce auxiliary fields. 
These auxiliary fields are  
important in picking out the correct zero modes {\it i.e.,} those which have 
the proper gauge and conformal properties, and can be eliminated as soon as 
this is done.   

Since the left and right-handed constraints are independent, it is easy to 
see that, in the light-cone version, the BFV reduction factor $F$ is just 
the product of two factors $F_\alpha$ and $F_{-\alpha}$ where 
$$ F_\alpha = \int dA^\alpha \delta (\chi^\alpha ) det\Bigl([FP]_{a^\alpha} +
 [BFV]_{a^\alpha}\partial_l\Bigr)~
 e^{\int dxdt ~[A^\alpha\Phi_\alpha ]}  
 \eqno(5.2)$$
and similarly for $F_{-\alpha}$. Furthermore, because $\Phi_\alpha =  
\pi_\alpha - {2k\over\alpha^2}M_\alpha$, we see that the argument in the
determinant in (5.2) is
$$[FP]_{a^\alpha} + [BFV]_{a^\alpha}\partial_l = -
\Bigl({\partial\chi^\alpha\over\partial a^\alpha} 
+ {\partial\chi^\alpha\over\partial A^\alpha}\partial_l\Bigr)\eqno(5.3)$$
According to the BFV prescription, the reduction factor (5.2) is to be 
inserted into the unconstrained WZW path integral (5.1). Thus, integrating 
over $\pi_i$ and regarding $\phi^i$ as a background field for the 
time being, the reduced path integral is 
$$I = \int d(a^{\pm\alpha}\pi_{\pm\alpha}A^{\pm\alpha}) 
 \delta (\chi^\alpha )
\delta (\chi^{-\alpha} )det\Bigl\lbrack{ 
\bigl({\partial\chi^\alpha\over\partial a^\alpha} 
+ {\partial\chi^\alpha\over\partial A^\alpha}\partial_l\bigr)
\bigl({\partial\chi^{-\alpha}\over\partial a^{-\alpha}} 
+ {\partial\chi^{-\alpha}\over\partial A^{-\alpha}}\partial_r\bigr)
\Bigr\rbrack}
~e^{-S_A} \eqno(5.4a)$$ 
where  
$$\eqalign{S_A = \int d^2z~[{k\over 2}(\partial_r\phi^i )&(\partial_l\phi^i ) 
+ \pi_\alpha\partial_l a^\alpha + \pi_{-\alpha}\partial_r a^{-\alpha} -
{\alpha^2\over 2k}\pi_\alpha\pi_{-\alpha} e^{\alpha\cdot\phi}\cr&-
A^\alpha (\pi_\alpha 
 - {2k\over\alpha^2}M_\alpha ) -
A^{-\alpha}(\pi_{-\alpha} - {2k\over\alpha^2}M_{-\alpha} )]} \eqno(5.4b)$$ 
Integrating over the momenta $\pi_\alpha$ and $ \pi_{-\alpha}$ gives the 
configuration space version of the BFV path integral for the gauged 
WZW model 
$$I = \int d(e^{-\alpha\cdot\phi}a^{\pm\alpha}A^{\pm\alpha} )
\delta (\chi^\alpha )
\delta (\chi^{-\alpha} )det\Bigl\lbrack{ 
\bigl({\partial\chi^\alpha\over\partial a^\alpha} 
+ {\partial\chi^\alpha\over\partial A^\alpha}\partial_l\bigr)
\bigl({\partial\chi^{-\alpha}\over\partial a^{-\alpha}} 
+ {\partial\chi^{-\alpha}\over\partial A^{-\alpha}}\partial_r\bigr)
\Bigr\rbrack}
~e^{-S_G} \eqno(5.5a)$$ 
where $S_G$ stands for the Action of the gauged WZW model and is given by 
$$S_G = \int d^2z~{k\over 2}(\partial_r\phi^i )(\partial_l\phi^i ) 
+ {2k\over\alpha^2}[e^{-\alpha\cdot\phi}(\partial_r a^{-\alpha} - 
A^{-\alpha})(\partial_l a^\alpha - A^\alpha ) 
+ A^\alpha M_\alpha + 
A^{-\alpha}M_{-\alpha}]  \eqno(5.5b)$$ 
Equations (5.5a,b) are the standard BFV results for the reduced path integral  
in Euclidean coordinates. It is obvious that the Action (5.5b) is invariant 
under the gauge transformations
$$a^\alpha \rightarrow a^\alpha + \lambda^\alpha , ~~~~~A^\alpha \rightarrow
 A^\alpha + 
\partial_l\lambda^\alpha \eqno(5.6)$$ 
and similarly for $a^{-\alpha}$ and $A^{-\alpha}$. 

We can now discuss the 
zero modes of the $A$'s. 
This we can do by taking into account the conformal 
spins $\omega (s_l, s_r)$ of the fields tabulated below  
$$\vbox{
\offinterlineskip
\halign{
\strut\vrule#&\vrule#&\vrule#&\vrule#&\vrule#\hfil\vrule\cr
\noalign{\hrule}
$~\omega (e^{\alpha\cdot\phi} )~$&$~\omega (a^\alpha )$ 
&$~\omega (a^{-\alpha} )$ &$~\omega (A^\alpha )~$&
$~\omega (A^{-\alpha})~$\cr
\noalign {\hrule}
$~(m_\alpha , m_\alpha )~$&$~(0, m_\alpha )~$&$~(m_\alpha , 0)~$&
$~(1, m_\alpha )~$&$~(m_\alpha , 1)~$\cr
\noalign{\hrule}
}}\eqno(5.7)$$
The weights of $a^\alpha , a^{-\alpha}$ and $e^{\alpha\cdot\phi}$
were determined following 
(3.25) and the natural choice of weights for the $A$ fields above follows 
from the gauge transformations (5.6).  
Consider, for definiteness,  $A^\alpha$, and decompose it into a maximally 
gauge invariant part $A^\alpha_0$ and its orthogonal complement $\hat A^\alpha$ 
which is gauge variant and can be gauged away {\it i.e.} let   
$$A^\alpha = A_0^\alpha + \hat A^\alpha~~{\hbox{where}}~~\hat A^\alpha = 
\partial_l\lambda^\alpha \eqno(5.8)$$
In the above equation the gauge transformation parameter $\lambda^\alpha$ 
has a conformal weight $\omega (\lambda^\alpha ) = (0, m_\alpha )$. 
The requirement that the $A^\alpha_0 $ and $\hat A^\alpha$ 
be orthogonal to each other can then be written as 
$$\int d^2z~h_\alpha A_0^\alpha\hat A^\alpha = 0
\eqno(5.9)$$
where $h_\alpha$ is a set of auxiliary fields with conformal weights 
$$\omega (h_\alpha ) = (-1, 1 - 2m_\alpha )
\eqno(5.10)$$
The factor $h_\alpha$ in (5.9) comes from the 
requirement that the orthogonality be defined in a conformally invariant 
manner. As is obvious from the weights of the $A$ fields defined in (5.7), 
it is not possible to define this condition in a conformally invariant manner 
without introducing auxiliary fields with appropriate conformal 
properties. It will become clear presently, however, that these fields are 
necessary only to pick  the true zero modes of the gauge fields. Once this is  
done, they can be easily eliminated.  
Substituting for $\hat A^\alpha$ from (5.8) in (5.9), and 
using the fact that the orthogonality must be valid for arbitrary 
$\lambda^\alpha$, it follows from a simple partial integration that 
$$\partial_l (h_\alpha A_0^\alpha ) = 0
~~~{\hbox{or}}~~~A_0^\alpha = h_\alpha^{-1}f^\alpha (z_r)\eqno(5.11)$$
where $f^\alpha (z_r)$ are arbitrary holomorphic functions with a conformal 
weights given by 
$$\omega (f^\alpha (z_r)) = (0, 1-m_\alpha )\eqno(5.12)$$
However, since there are no holomorphic (or anti-holomorphic) functions  
on a compact Riemann surface except the constant functions [11], we see  
that $f^\alpha (z_r)$ must be constant. Furthermore, for $m_\alpha \neq 1$,  
$f^\alpha (z_r)$ are not conformal scalars and hence the only constant they  
can be set equal to, without breaking conformal invariance, is zero. 
Thus there are no zero modes which have grade greater than one and only one 
zero mode of grade one for each positive root, namely,  
$$A^{\alpha^{(s)}}_0 \sim  h_{\alpha^{(s)}}^{-1}\eqno(5.13)$$ 
We may now eliminate the auxiliary field by setting 
$$ h_{\alpha^{(s)}} = e^{\alpha^{(s)}\cdot\phi}
\eqno(5.14)$$
This is the most natural choice for $h_\alpha$ 
since $e^{\alpha^{(s)}\cdot\phi}$ are the only local fields, apart from 
the background metric,  which 
has the correct conformal weight.

A more intuitive argument would be to 
note that any gauge-invariant part $A^\alpha_0$ of $A^\alpha$ would have 
conformal spin (the difference of the left and right conformal weights)  
equal to $1-m_\alpha$ and   
be both primary and local. 
But the only residual fields out of which they could be constructed are the 
fields $e^{\alpha\cdot\phi}$ and since these fields have spin zero,  
the only permissible zero modes are those that have spin zero {\it i.e}  
grade equal to one.  
The Lagrange multiplier fields can therefore be written as  
$$A^\alpha  = \mu^\alpha e^{\alpha\cdot\phi} + \partial_l\lambda^\alpha ,~~~
A^{-\alpha} = \mu^{-\alpha} e^{\alpha\cdot\phi} +
 \partial_r\lambda^{-\alpha}~~~{\hbox{where}}~~~\mu^\alpha , \mu^{-\alpha}
\neq 0~{\hbox {iff}}~\alpha\in\Delta_s \eqno(5.15)$$
$\mu^\alpha$ and $\mu^{-\alpha}$ being constants.

As has already been mentioned, it is desirable to have a curved space
generalisation of the path integral in (5.5) because the manifold is 
compact. We choose  
a fixed background metric $g^{\mu\nu}$ for this purpose.   
An interesting property of the Action (5.5b) is that, if we use conformal  
coordinates $g_{\mu\nu} = e^{\sigma (x)}\eta_{\mu\nu}$,  
the metric does not appear explicitly.   
In particular, since the partial derivatives 
act on the sides of the fields that have conformal weight zero, they 
remain ordinary derivatives {\it i.e.} there is no need to modify them with  
the spin connection $\partial\sigma$. 
Furthermore, this continues to be the case when we change 
from the Sugawara Virasoro to the improved one. 
The reason for the invariance under the 
change of Virasoro is that  
the change of $a^\alpha$ and $a^{-\alpha}$ from scalars 
to fields of weights $(0, m_\alpha )$ and $(m_\alpha , 0)$  
respectively is exactly compensated by the change in $e^{\alpha\cdot\phi}$ from 
a conformal scalar to a primary field of weight $(m_\alpha , m_\alpha )$. 

As mentioned earlier, the improvement terms in the Virasoro can be incorporated 
explicitly in the presence of a background metric. This is done by  
adding to the 
Lagrangian density, a term of the form 
$\sqrt g g^{\mu\nu} \nabla_\mu J_\nu^0$, where $\nabla_\mu = \partial_\mu 
+ (\partial_\mu\sigma)$ is the covariant derivative.  
In conformal coordinates this 
reduces to $(\partial_\mu\sigma )J_\mu^0$ 
apart from a total derivative term. However, since  
the field $\alpha\cdot\phi $ 
is no longer a scalar but a spin-zero connection, the current $J_\mu^0$ 
is no longer a vector but a spin-one connection. To restore the vectorial 
properties of $J_\mu^0$, it is necessary to let $J_\mu^0\rightarrow 
J_\mu^0 - \rho\partial_\mu\sigma$. In that case, the cross-terms in 
$ Tr~ (J^i)^2  
+ (\partial_\mu\sigma )J_\mu^0$ exactly cancel leaving a net addition  
to the Lagrangian density of a Polyakov term $-k\rho^2 (\partial\sigma )^2/2$. 
The Polyakov term cannnot be ignored 
because it is this term that produces the known classical centre 
$c = -k\rho^2$ for the improved Virasoro algebra according to the standard 
formula $\partial S/ \partial\sigma (x) = cR(x)$, where $R(x)$ is the Ricci 
scalar. Thus the net effect of introducing curvilinear coordinates is simply to 
add a Polyakov term $k\rho^2 R\sigma$/2 to the Action.  

We also have to consider 
the effect of the change of Virasoro on the 
measure in (5.5). The factor    
$(e^{-\alpha\cdot\phi}a^\alpha a^{-\alpha} )$ in the measure remains a scalar 
under the change of Virasoro. Hence the curved space generalisation of the 
$a^\alpha a^{-\alpha}$ integral 
requires only the usual factor $\sqrt g$. On the other hand, since the 
$A^\alpha$ and $A^{-\alpha}$ fields have weights $(1, m_\alpha )$ and  
$(m_\alpha , 1)$ respectively, their measure requires a factor
$({1\over \sqrt g})^{m_\alpha}$.    

Substituting (5.15) 
in the gauged WZW path integral (5.5a,b), and incorporating the above 
mentioned modifications because of the curved space generalisation, we get
$$\eqalign{I = \int d(\sqrt g &e^{-\alpha\cdot\phi}a^{\pm\alpha} )
d\Bigl\lbrack ({1\over\sqrt g})^m 
\partial_l\lambda^\alpha\partial_r\lambda^{-\alpha}
\Bigr\rbrack\cr &\delta (\chi^\alpha )
\delta (\chi^{-\alpha} )
det\Bigl\lbrack{ 
\bigl({\partial\chi^\alpha\over\partial a^\alpha} 
+ {\partial\chi^\alpha\over\partial \lambda^\alpha}\bigr)
\bigl({\partial\chi^{-\alpha}\over\partial a^{-\alpha}} 
+ {\partial\chi^{-\alpha}\over\partial\lambda^{-\alpha}}\bigr)\Bigr\rbrack}
~e^{-{\hat S}_G}\times I_0 }\eqno(5.16a)$$ 
where ${\hat S}_G $ is the Action for the fluctuations and $I_0$ is the path 
integral for the zero modes. Since the cross-terms between 
the $A^\alpha_0$ and $\hat A^\alpha$ terms, as well as the 
$M_\alpha\hat A^\alpha$ and $M_{-\alpha}\hat A^{-\alpha}$ 
terms are pure divergences,  these terms drop out and ${\hat S}_G$ and $I_0$ 
may be written as   
$${\hat S}_G = \int d^2z~[{k\over 2}(\partial_r\phi^i )(\partial_l\phi^i ) 
+ {2k\over\alpha^2} e^{-\alpha\cdot\phi}\Bigl(\partial_l
(a^\alpha - \lambda^\alpha)\Bigr)
\Bigl(\partial_r (a^{-\alpha} - \lambda^{-\alpha})\Bigr) 
- {k\rho^2\over 2}(\partial\sigma )^2] \eqno(5.16b)$$ 
and 
$$I_0 = \int d(\mu^\alpha\mu^{-\alpha})~e^{-\int d^2z~[
{2k\over\alpha^2}e^{\alpha\cdot\phi}
\mu^\alpha\mu^{-\alpha} - {2k\over\alpha^2}e^{\alpha\cdot\phi}  
(\mu^\alpha M_\alpha + \mu^{-\alpha}M_{-\alpha})] } = 
e^{k\Lambda_\alpha\int d^2z~e^{\alpha\cdot\phi}}\eqno(5.16c)$$
respectively. Note that the integral over the zero modes $\mu^{\pm\alpha}$ has
produced the Toda potential term  
$k\Lambda_\alpha e^{\alpha\cdot\phi},~~\Lambda_\alpha \neq 0,~~ 
{\hbox {iff}}~ \alpha\in\Delta_s$. The determinant in (5.16a) 
may be simplified  
by using 
$${\partial\chi^\alpha\over\partial a^\alpha } + {\partial\chi^\alpha
\over \partial 
\lambda^\alpha} = 2{\partial\chi^\alpha\over\partial 
(a^\alpha + \lambda^\alpha )}\eqno(5.17)$$
and a similar expression for  
$\chi^{-\alpha}$ and $a^{-\alpha}$. Upon using this result, the  measure in 
(5.16a) reduces to 
$$ 4d(\sqrt g e^{-\alpha\cdot\phi} a^\alpha a^{-\alpha} )
d\Bigl\lbrack ({1\over\sqrt g})^{m_\alpha} \partial_l\lambda^\alpha\partial_r
\lambda^\alpha \Bigr\rbrack
\delta (a^\alpha + \lambda^\alpha )
\delta (a^{-\alpha} + \lambda^{-\alpha} )\eqno(5.18)$$ 
Eliminating the $\lambda$ fields by means of the delta functions and rescaling 
$a^\alpha$ and $a^{-\alpha}$ by a factor 2, the path integral becomes 
$$I = I_\alpha\times 
det \Bigl\lbrack ({1\over\sqrt g})^{m_\alpha} \partial_r\partial_l\Bigr\rbrack
~e^{-\int d^2z~[{k\over 2}(\partial_r\phi^i )(\partial_l\phi^i ) 
- k\sum_{\alpha\in\Delta_s}\Lambda_\alpha e^{\alpha\cdot\phi} 
-{k\rho^2\over 2}(\partial\sigma )^2]}\eqno(5.19)$$
The $I_\alpha$ in (5.19) stands for the $a^\alpha a^{-\alpha}$ part of the
integral and is just the well-known one 
encountered in the computation of the WZW partition function, namely, 
$$\eqalign{I_\alpha &= \int d(\sqrt g e^{-\alpha\cdot\phi}
a^{\pm\alpha} ) 
e^{-k\int d^2z~e^{-\alpha\cdot\phi}(\partial_l a^\alpha )
(\partial_r a^{-\alpha} )}\cr& = 
\int d(ad)~e^{-\int d^2z~a\Bigl(g^{-{1\over 4}}(D_l^{\alpha\cdot\phi})^T
(D_r^{\alpha\cdot\phi})
g^{-{1\over 4}}\Bigr)d}  \cr
&= det \Bigl({1\over \sqrt g}(D_l^{\alpha\cdot\phi})^T
(D_r^{\alpha\cdot\phi})\Bigr)^{-1}}\eqno(5.20)$$ 
where $a = \sqrt k g^{1\over 4}e^{-{\alpha\cdot\phi\over 2}} a^\alpha$,   
$d = \sqrt k g^{1\over 4}e^{-{\alpha\cdot\phi\over 2}}a^{-\alpha}$,   
$D^{\alpha\cdot\phi} = \partial + (\partial\alpha\cdot\phi )$ and 
$(D^{\alpha\cdot\phi} )^T = \partial - (\partial\alpha\cdot\phi )$. 
Thus the path integral (5.19)  may be expressed as 
$$I = e^{-k\int d^2z~[{1\over 2}(\partial_r\phi^i )(\partial_l\phi^i ) 
- \sum_{\alpha\in\Delta_s}\Lambda_\alpha e^{\alpha\cdot\phi} - 
{1\over 2}(\partial\sigma )^2]}
{det\Bigl\lbrack ({1\over\sqrt g})^{m_\alpha}(\partial_l\partial_r)\Bigr\rbrack
\over det \Bigl\lbrack {1\over \sqrt g}
(D_l^{\alpha\cdot\phi} )^T(D_r^{\alpha\cdot\phi} )\Bigr\rbrack}\eqno(5.21)$$ 
However, from the identity [12], 
$$ det \Bigl\lbrack ({1\over\sqrt g})^{m_\alpha}
 (\partial_l\partial_r)\Bigr\rbrack = 
det \Bigl\lbrack ({1\over\sqrt g}) (\partial_l\partial_r)\Bigr\rbrack
\times  e^{{Q_\alpha\over 2}\int R{1\over\nabla }R}
~~~{\hbox{where}}~~~Q_\alpha  = m_\alpha (m_\alpha -1)\eqno(5.22)$$
and the well-known WZW anomaly equation [13]  
$${det {1\over\sqrt g}(\partial_l\partial_r)\over det {1\over \sqrt g}
(D_l^{\alpha\cdot\phi} )^T(D_r^{\alpha\cdot\phi} )} = 
e^{{1\over 2}\int d^2z~\sum_\alpha [ 
(\partial\alpha\cdot\phi )^2 
- \sqrt gR\alpha\cdot\phi ]}
= e^{\int d^2z~ [ 
{1\over 2}(\partial_r\alpha\cdot\phi^i)(\partial_l\alpha\cdot\phi^i) 
- \sqrt gR\rho\cdot\phi ]}
\eqno(5.23)$$ 
we obtain 
$$I = e^{-\int d^2z 
[{(k - \gamma )\over 2}(\partial_r\phi^i )(\partial_l\phi^i ) -  
k\sum_{\alpha\in\Delta_s}\Lambda_\alpha e^{\alpha\cdot\phi} +
 \sqrt g (j +  R\rho )\cdot\phi - 
{k - (\gamma - 2)\over 2}\rho^2 (\partial\sigma )^2]}
\eqno(5.24)$$
where we have used $m_\alpha = \rho\cdot\alpha$ 
and $\sum\alpha_i\alpha_i = \gamma\delta_{ij}$, $\gamma$ being 
the dual Coxeter number, to get   
$\sum Q_\alpha =  (\gamma - 2)\rho^2$.  
Reintroducing the $\phi$-integration we have finally the 
reduced configuration space path integral for $\phi$ 
$$I = \int d(g^{1\over 4}\phi^i )~e^{-\int d^2z 
[{(k - \gamma )\over 2}(\partial_r\phi^i )(\partial_l\phi^i ) -  
k\sum_{\alpha\in\Delta_s}\Lambda_\alpha e^{\alpha\cdot\phi} +
 \sqrt g (j +  R\rho )\cdot\phi - 
{k - (\gamma - 2) \over 2}\rho^2 (\partial\sigma )^2]}
\eqno(5.25)$$
In the flat space limit this reduces to 
$$I = \int d\phi^i ~e^{-\int d^2z 
[{(k - \gamma )\over 2}(\partial_r\phi^i )(\partial_l\phi^i ) -  
k\sum_{\alpha\in\Delta_s}\Lambda_\alpha e^{\alpha\cdot\phi} + j\cdot\phi ]} 
\eqno(5.26)$$
which is easily recognised as the path integral for the Toda theory. Thus  
(5.25) is just the Toda path integral in a fixed curved background. Writing 
$\phi^i =  s^i  - \rho^i\sigma $, so that $s^i$ is a 
scalar field, (5.24) becomes 
$$I = \int d(g^{1\over 4}\phi )~e^{-\int d^2z 
\bigl[{(k - \gamma )\over 2}(\partial_r s^i )(\partial_l s^i ) 
- k\sum_{\alpha\in\Delta_s}\Lambda_\alpha\sqrt g e^{\alpha\cdot s} +
 \sqrt g \bigl ( 1 + (k - \gamma )\bigr) R\rho\cdot s 
+ \sqrt g j\cdot s\bigr]}
\eqno(5.27)$$
where the terms that depend purely on $\sigma$, including the Polyakov term,
 have cancelled (except for 
the $j\sigma$ term which we have dropped). It is well-known that the 
Virasoro centre for this theory has the form 
$$c(k - \gamma ) =  \hbar\Bigl\lbrack {l\over 6} + {2\rho^2\over (k - \gamma )}
\bigl[ 1 +
 (k - \gamma )\bigr]^2\Bigr\rbrack =   
l {\hbar\over 6} + 
{2\rho^2\over (\kappa - \gamma\hbar )}\Bigl[ \hbar +
 (\kappa - \gamma\hbar )\Bigr]^2\eqno(5.28)$$  
where $l = {\hbox {dim}}H$.   
The $\hbar$/6 in (5.28) comes from the Weyl anomaly for a
single scalar field 
and, to separate the quantum effects, we have recalled from (3.1) that 
$k = \kappa /\hbar$ where $\kappa \sim 1$. 
The results are independent of the choice of the gauge 
fixing conditions -- as predicted by the Fradkin-Vilkovisky theorem. 

Note that the WZW anomaly (ratio of determinants) was produced by the 
fact that the integration over the $\hat A$ fields is restricted by the 
condition that $\hat A$ be a gradient. 
Had $A$ been free, we could have integrated 
directly over $A$ in (5.5) and there would have been no WZW anomaly (although 
there would still be a Weyl anomaly). Thus the WZW  
anomaly, originates in the fact that the gauge variant parts of the 
Lagrange multipliers are gradients.  
The presence of the WZW anomaly means that although the classical reduction 
converts the WZW theory into a Toda  theory with coupling constant $k$,
the quantum reduction converts it into a Toda  theory with coupling 
constant $k - \gamma$. As a result, although the expressions for the 
Virasoro centre in (3.26) and (5.28) have the same functional form, their 
arguments are $k$ and  $k - \gamma$ respectively. Thus the operations of 
reducing and quantising do not commute.   
\bigskip
\centerline {\bf VI. THE TODA AND WZW GAUGES} 
\bigskip
We would now like to investigate what happens in two particular gauges namely, 
the Toda (physical) gauge and the WZW gauge. The Toda gauge 
highlights the origin of the WZW anomaly. 
The WZW gauge 
provides an interesting interpretation of the formula (5.28) for the 
Virasoro centre.
The key equation  
for comparison is (5.5), just prior to the separation of the $A$ fields
into their zero mode and gauge variant parts.

The Toda gauge is defined by   
$\chi^\alpha  \equiv a^\alpha$. 
In this gauge, one sees 
from (4.7) that $[FP] = -1$ and $[BFV] = 0$. Hence the $a^{\pm\alpha}$ 
fields are eliminated by the delta functions and we obtain 
$$I = \int det (e^{-\alpha\cdot\phi}) d(A^{\pm\alpha} )
~e^{- 
\int d^2z~{k\over 2}(\partial_r\phi^i )(\partial_l\phi^i ) 
+ {2k\over\alpha^2}[e^{-\alpha\cdot\phi} 
A^{-\alpha} A^\alpha  
+ A^\alpha M_\alpha + 
A^{-\alpha}M_{-\alpha}]}  \eqno(6.1)$$ 
which on separating the zero modes and integrating over them as in (5.16c)
becomes 
$$I = \int det (e^{-\alpha\cdot\phi}) d(\hat A^{\pm\alpha} )~e^{- 
\int d^2z~{k\over 2}(\partial_r\phi^i )(\partial_l\phi^i ) + k\sum_{\alpha\in
\Delta_s}\Lambda_\alpha  
e^{\alpha\cdot\phi} + {2k\over\alpha^2}e^{-\alpha\cdot\phi} 
\hat A^{-\alpha} \hat A^\alpha  } 
 \eqno(6.2)$$ 
Note that if there were no zero modes, the $M$-dependent 
terms would drop out and there would be no Toda potential. 

We now use the fact that the $\hat A$ fields are gradients, to obtain  
$$\eqalign{det &(e^{-\alpha\cdot\phi}) d\hat A^{\pm \alpha} =
 det (e^{-\alpha\cdot\phi})
d(\partial_l\lambda^\alpha \partial_r\lambda^{-\alpha}) \cr& = det  
(e^{-\alpha\cdot\phi})d\lambda^{\pm\alpha} det (\partial_r\partial_l ) = d(e^{ 
-\alpha\cdot\phi}\lambda^{\pm\alpha}) det (\partial_r\partial_l )}\eqno(6.3)$$
and the rest of the integration proceeds as before. The  
important point to note is that had the $\hat A$ fields not been gradients, we 
could have replaced 
$ det (e^{-\alpha\cdot\phi})d\hat A^{\pm\alpha}$ by  
$d(e^{-\alpha\cdot\phi}\hat A^{\pm\alpha})$ and then there would have been 
no anomaly. But because $\hat A$ is a gradient, we have  
$$\eqalign{d&(e^{-\alpha\cdot\phi}\hat A^{\pm\alpha}) = d(e^{-\alpha\cdot\phi}
\partial_l\lambda^\alpha\partial_r\lambda^{-\alpha})  
\cr &= d\Bigl\lbrack (D_l^{\alpha\cdot\phi})^T(D_r^{\alpha\cdot\phi}) 
 e^{-\alpha\cdot\phi}
\lambda^{\pm\alpha}\Bigr\rbrack =
det \Bigl\lbrack (D_l^{\alpha\cdot\phi})^T(D_r^{\alpha\cdot\phi})\Bigr\rbrack
d(e^{-\alpha\cdot\phi}\lambda^{\pm\alpha})}\eqno(6.4)$$ 
which is not the same as the correct measure (6.3). 
Thus the Toda gauge  shows explicitly how the zero modes 
produce the Toda potential and the gauge variant modes produce the  
WZW anomaly.  

The WZW gauge is defined by $\chi^\alpha \equiv \hat A^\alpha$. In this gauge,  
$\hat A^\alpha = 0$, and one sees from (4.7) that $[FP] = 0$ and $[BFV] = -1$,
which, incidentally, shows the necessity of using the BFV formalism for 
considering this gauge. In this case, the $\hat A$ fields are eliminated 
by the gauge fixing delta functions and on integrating over the zero modes of 
the $A$ fields, the reduced path integral (5.5) becomes   
$$\eqalign{I = \int& d(e^{-\alpha\cdot\phi}a^{\pm\alpha} )\times 
\cr&e^{- 
\int d^2z~\Bigl[ {k\over 2}(\partial_r\phi^i )(\partial_l\phi^i ) 
+ ({2k\over\alpha^2}) e^{-\alpha\cdot\phi}(\partial_r a^{-\alpha}  
)(\partial_l a^\alpha ) + k\sum_{\alpha\in\Delta_s}\Lambda_\alpha
 e^{\alpha\cdot\phi} + {(k - (\gamma - 2))\rho^2\over 2}R\sigma 
\Bigr]}}\eqno(6.6)$$ 
Note that the Action in (6.6) is just the original WZW Action together with 
the exponential term and the Polyakov term. This form of the Action allows 
us to read off the Virasoro centre by inspection namely, 
$$\eqalign{{6c\over\hbar} &=  {k r\over k - \gamma} + 12\rho^2k - 
 (r - l + 12 \sum Q_\alpha ) = l +  
{\gamma r\over k - \gamma} + 12\rho^2 (k -\gamma + 2)\cr&= 
l + 24\rho^2 + 12\rho^2[{1\over k - \gamma} + k - \gamma ]}\eqno(6.7)$$
where $r = {\hbox {dim}}G$ and $l = {\hbox {dim}}H$. 
For the last equality above we have used the `strange' formula 
$ 12\rho^2 = \gamma r$ of Freundenthal and deVries [14].   
The first term in (6.7) comes from the WZW piece and the second term from 
the Polyakov term which has contributions from the classical 
improvement to the Virasoro and from the ghosts. The combination of these 
terms simplifies to (5.28) and     
thus gives a simple interpretation of the Toda centre as the sum of the  
WZW centre and the classical improvement centre $k$, minus the ghost centre
$r-l + \sum Q_\alpha$. The expression (6.7) for the centre was obtained 
earlier in [1] without using the curved background. 
\bigskip
\centerline {\bf VII. CONCLUSIONS}
\bigskip
We have shown that the quantum mechanical WZW $\rightarrow$ Toda reductions 
can be formulated by means of the path integral in a gauge independent manner.
For this purpose we have used a 
modification of the 
conventional Batalin, Fradkin,
Vilkovisky formalism for first class constraints which takes into account 
the chirality of the constraints and the compactness of the manifold.  
An interesting 
feature of the reduction is the role played by the decomposition of the 
Lagrange multipliers into zero modes which are gauge invariant and 
gradient parts which are not. The zero modes  produce the Toda potential and 
the gradients produce the WZW anomaly $(k\rightarrow k - \gamma )$.
This anomaly plays a crucial role   
in proving the Fradkin-Vilkovisky theorem regarding  
gauge invariance of the reduction. This is shown explicitly by the 
fact that if the anomaly is neglected, the centre of the Virasoro algebra  
is $c(k)$ and $c(k - \gamma )$ in the Toda and WZW gauges respectively. 

Another interesting feature is that the operations of reducing and 
quantising do not commute in the sense that they lead to Toda 
theories with different coupling constants. Reduction of the classical 
WZW theory leads 
to a Toda coupling constant $k$ -- which is not changed by 
subsequent quantisation -- whereas reduction of the quantised 
WZW theory leads, as we have seen,  
to a Toda coupling constant $k -\gamma$.  

The reduction was simplified by the fact that, in conformal coordinates,
the change from the 
Sugawara to the `improved' energy momentum tensor (which is necessary for 
conformal invariance) does not show up explicitly in the path integral except  
for Polyakov terms which appear when the fields are coupled to a fixed 
background metric. This means that for flat (toroidal) spaces, the path 
integral remains form invariant under the change of Virasoro. 
The basic reason for this is that although the 
individual fields change their conformal properties with respect to the 
improvement, the combination 
$e^{-\alpha\cdot\phi}a^\alpha a^{-\alpha}$ which appears in both the measure 
and the Lagrangian, is invariant under the change. 

The modification of the BFV formalism that we have used should be useful 
for other conformal reductions such as WZW $\rightarrow$ non-abelian Toda
reductions and the coset constructions of Goddard and Olive [15]. We hope 
to address these questions in the future.  
\vfil\eject
\centerline {\bf {REFERENCES}}
\bigskip
\item {1. } F. A. Bais, T. Tjin and P. Van Driel, Nuc. Phys. {B 357} (1991) 
632; V. A. Fateev and S. L. Lukyanov, Int. J. Mod. Phys. {A3} (1988) 507;
S. L. Lukyanov, Funct. Anal. Appl. {22} (1989) 255; P. Forg\'acs, A. Wipf,
J. Balog, L. Feh\'er, and L. O'Raifeartaigh, Phys. Lett. {B227} (1989)214;
J. Balog, L. Feh\'er, L. O'Raifeartaigh, P. Forg\'acs and A. Wipf, Ann. Phys. 
{203} (1990) 76; Phys. Lett {B244}(1990)435; L. Feh\'er, 
L. O'Raifaertaigh, P. Ruelle, I. Tsutsui and A. Wipf, Phys. Rep. { 222}
No. 1, (1992)1; P. Bouwknegt and K. Schoutens Eds. W Symmetry (Advanced 
series in mathematical physics, 22) World Scientific, Singapore (1995).  

\item {2. } A. Bilal and J. L. Gervais, Phys. Lett {B206} (1988) 412;
Nuc. Phys. { B314} (1989) 646; { B318} (1989) 579; O. Babelon, Phys. 
Lett. { B215} (1988) 523, T. Hollowod and P. Mansfield, Nuc. Phys. 
{ B330} (1990) 720; P. Mansfield and B. Spence, Nuc. Phys. { B362}  
(1991) 294; P. Bowcock and G. M. T. Watts, Nuc. Phys. { B379} (1992) 63;
C. Ford and L. O'Raifeartaigh, Nuc. Phys. { B460} (1996) 203. 

\item {3. } L. O'Raifeartaigh and V. V. Sreedhar, Path Integral Formulation 
of the Conformal Wess-Zumino-Witten $\rightarrow$ Liouville Reduction, 
Phys. Lett. B, {\it in press}. 

\item {4. } I. A. Batalin and E. S. Fradkin, in: Group Theoretical Methods 
in Physics, Vol II (Moscow, 1980); I. A. Batalin and G. Vilkovisky, 
Phys. Lett. { B69} (1977) 309; E. S. Fradkin and G. A. Vilkovisky, 
Phys. Lett. { B55} (1975) 224; M. Henneaux, Phys. Rep { 126} No.1, 
(1985) 1. 

\item {5. } A. O. Barut and R. Raczka, Theory of Group Representations 
and Applications; PWN Polish Scientific Publishers, Warszawa 1977. 

\item {6. } E. Witten, Comm. Math. Phys. { 92} (1984) 483; P. Goddard and 
D. Olive, Int. J. Mod. Phys. { A1} (1986) 303; P. Di Francesco, P. Mathieu,
and D. S\'en\'echal, Conformal Field Theory, Graduate Texts in Contemporary 
Physics, (Springer-Verlag New York, Inc. 1997).     

\item {7. } I. Tsutsui and L. Feh\'er, Prog. Theor. Phys. Suppl. { 118} 
(1995), 173. 

\item {8. } A. M. Polyakov and P. B. Wiegmann, Phys. Lett B 131 (1983) 121. 

\item {9. } J. Govaerts, Int. J. Mod. Phys. A4 (1989) 173, Int. J. Mod. Phys. 
A4 (1989) 4487, Class. Quant. Grav. 8 (1991) 4487.  

\item {10. } L. D. Faddeev, Theor. Math. Phys. { 1} (1970) 1;
 L. D.  Faddeev and  A. A. Slavnov, Gauge Fields, introduction 
to quantum theory. (Benjamin Cummins, Reading, Massachusetts, 1980).  

\item {11. } Phillip A. Griffiths, Introduction to Algebraic Curves, 
Vol. { 76} Translations of Mathematical Monographs, American 
Mathematical Society (1989). 

\item {12. } M. B. Green, L. H. Schwarz and E. Witten, Superstring Theory,
Vol. 1, Cambridge University Press 1987. 

\item {13. } M. L\"uscher, Ann. Phys. 142 (1982) 359; H. Leutwyler and 
S. Mallik, Z. Phys. C33 (1986) 205.  

\item {14. } H. Freudenthal and H. de Vries, Linear Lie Groups, Academic 
Press, New York and London, 1969. 

\item {15. } P. Goddard and D. I. Olive, Int. J. Mod. Phys. A1 (1983) 303.
\vfil\eject\end